\documentclass[aps,prl,preprint,superscriptaddress]{revtex4-1}

\usepackage{graphicx}
\usepackage{amssymb, amsmath}
\usepackage[usenames]{color}
\usepackage{dcolumn}
\usepackage{bm}
\usepackage{ textcomp }

\begin{document}


\title{Alloy engineering of topological semimetal phase transition in MgTa$_{2-x}$Nb$_x$N$_3$}


\author{Huaqing Huang}
\affiliation{Department of Materials Science and Engineering, University of Utah, Salt Lake City, Utah 84112, USA}

\author{Kyung-Hwan Jin}
\affiliation{Department of Materials Science and Engineering, University of Utah, Salt Lake City, Utah 84112, USA}

\author{Feng Liu\footnote{Corresponding author: fliu@eng.utah.edu}}
\affiliation{Department of Materials Science and Engineering, University of Utah, Salt Lake City, Utah 84112, USA}
\affiliation{Collaborative Innovation Center of Quantum Matter, Beijing 100084, China}

\date{\today}

\begin{abstract}
Dirac, triple-point and Weyl fermions represent three topological semimetal phases, characterized with a descending degree of band degeneracy, which have been realized separately in specific crystalline materials with different lattice symmetries. Here we demonstrate an alloy engineering approach to realize all three types of fermions in one single material system of MgTa$_{2-x}$Nb$_x$N$_3$. Based on symmetry analysis and first-principles calculations, we map out a phase diagram of topological order in the parameter space of alloy concentration and crystalline symmetry, where the intrinsic MgTa$_2$N$_3$ with the highest symmetry hosts the Dirac semimetal phase which transforms into the triple-point and then the Weyl semimetal phase with the increasing Nb concentration that lowers the crystalline symmetries. Therefore, alloy engineering affords a unique approach for experimental investigation of topological transitions of semimetallic phases manifesting different fermionic behaviors.
\end{abstract}

\pacs{}

\maketitle

Alloying engineering is one of the most well-established approaches to tailor the material's structural, mechanical and electronic properties \cite{pfeiler2008alloy,chen2012semiconductor}. For example, the classical Vigard's law underlies that the lattice constant of an alloy A$_{1-x}$B$_x$ is usually a linear interpolation of the lattice constants of A and B \cite{vegard1921konstitution,PhysRevA.43.3161}. The well-known bowing curve provides a useful guideline for determining the band gap of a semiconductor alloy \cite{PhysRevB.1.3351, PhysRevLett.51.662, richardson1972origins}. Recently, alloy engineering has also been extended to tailor the 
topological material properties \cite{hsieh2008topological, Na3BiSb, HgTeS, PbSnTe,HsinMoWTe2,*HasanMoWTe2,huanghqIIIV}, such as the Fermi energy, spin chirality, Fermi arcs and the phase transition between trivial and topological states. In this Letter, we demonstrate for the first time that alloy engineering can also be applied to tune the transitions among different topological fermions in topological semimetals. 

Topologically protected fermionic quasiparticles in semimetals are directly related to the degeneracy of band crossing points around the Fermi level which are determined by the cyrstalline symmetry.
Fourfold Dirac fermions in Dirac semimetals, such as Na$_3$Bi \cite{Na3Bi,*Na3Bi_exp1,*Na3Bi_exp2,*Na3Bi_exp3} and Cd$_3$As$_2$ \cite{Cd3As2,*Cd3As2_exp1,*Cd3As2_exp2,*Cd3As2_exp3}, are protected by inversion, time-reversal and rotational symmetries.
Weyl fermions with twofold degeneracy are observed in Weyl semimetals with broken inversion \cite{TaAsHLin,*xu2015discovery,*BinghaiTaAs2015, TaAsPRX,*PhysRevX.5.031013,tripleBinghai} or time-reversal symmetry \cite{ZJWangTRbreaking}. Triple-point fermions, an intermediate state between fourfold Dirac and twofold Weyl fermions, can be realized in noncentrosymmetric WC-type materials with certain crystalline symmetries \cite{triple1,triple2,tripleNature,*ma2017three}. Although different fermions are realized in different crystal materials with specific lattice symmetry, few of them can exist in a single material simultaneously \cite{sunjianPRL,sunjianArxiv,triple3,triple4}. It is well known that a Dirac fermion would split into a pair of Weyl fermions by breaking either inversion or time-reversal symmetry of a Dirac semimetal \cite{Na3Bi, PhysRevB.95.161306, PhysRevB.88.165105}. Actually, it is theoretically possible to realize all these fermions by gradually reducing crystalline symmetry of certain compounds. However, such three-stage transition in a real material has never been seen yet, also the study of coexistence and interplay between these fermions is still lacking. Hence realizing the transition between Dirac, triple-point and Weyl fermionic quasiparticles in one single material system is of great importance for material science, condensed matter physics and elementary particle physics.

Here, we propose an alloy engineering approach to realize all three types of fermions in a single material system. Taking MgTa$_{2-x}$Nb$_x$N$_3$ as en example, we show that Dirac fermions which exist in the intrinsic MgTa$_2$N$_3$ with the highest symmetry can be converted into triple-point and Weyl fermions with the increasing Nb concentration that lowers the cyrstalline symmetries. The evolution of Fermi arcs and quasiparticle interference (QPI) spectra in different topological semimetallic states have been studied in depth. Our findings not only are of fundamental interest for a better understanding of basic properties of different topological fermions but also provide a unique approach for experimental investigation of phase transitions among different topological states.

\begin{figure}
\includegraphics[width =1\columnwidth]{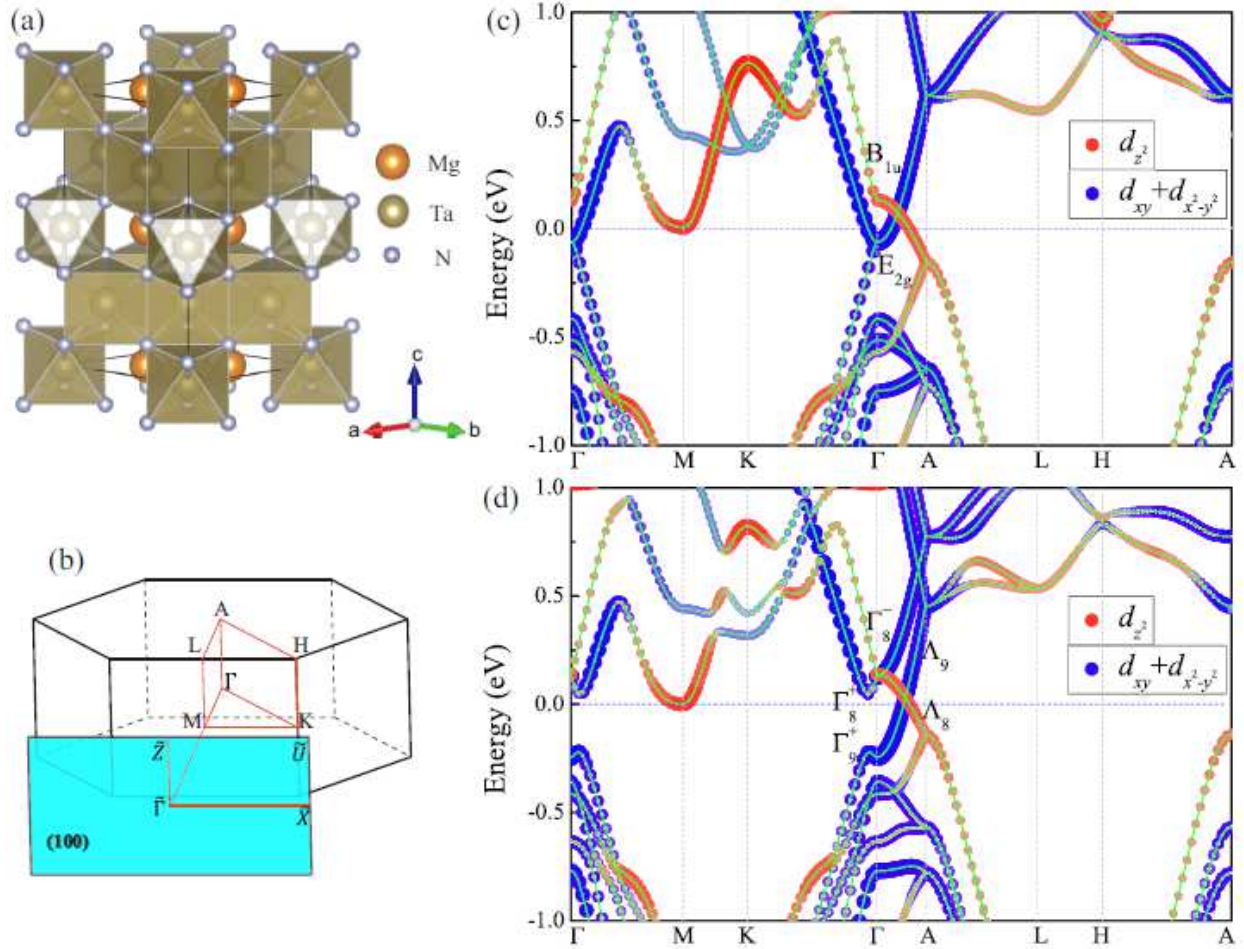}
\caption{\label{fig1_stru} (Color online) (a) Crystal structure of MgTa$_2$N$_3$ with $P6_3/mcm$ (No. 193) symmetry in a hexagonal cell. (b) The Brillouin zone of MgTa$_2$N$_3$ and the projected surface Brillouin zones of the (100) surface. Band structure of MgTa$_2$N$_3$ (c) without SOC and (d) with SOC.
Red and blue dots in band structures indicate the projection onto the Ta $d_{z^2}$ and $d_{xy}+d_{x^2-y^2}$ orbitals, respectively.}
\end{figure}

We performed first-principles calculations for MgTa$_{2-x}$Nb$_x$N$_3$ using the Vienna \textit{ab initio} simulation package \cite{VASP}.
Details of computational methods are presented in the Supplemental Materials \footnote{\label{fn}See Supplemental Material at http://link.aps.org/supplemental/xxx, for more details about the computation, which include Refs.~\cite{PBE,PAW,wannier90,huanghqAg2S,lopez,*lopez2,wu2017wanniertools,slater1964atomic,clementi1967atomic,pyykko2009single,pyykko2005triple,
cordero2008covalent,handbook2017,welham1999ambient,jacobs1989synthese,rauch1992ambient,jacobs1993synthesis,jacobs1993synthesis,
ohkubo2015anisotropic,balbarin1996high,jacobs1993synthesis,zakutayev2014experimental,zachwieja1991cutan,miura2011silver,niewa2004metal,
bowman2003synthesis,chen1998synthesis,helmlinger1993ba2,seeger1994synthesis,cario2001ln,chen1997synthesis,chen1994new,chen1994synthesis,
yang2015weyl,qing2012interface,tan2013interface,tu2017large,PhysRevLett.95.146802,huanghqInterface}.}. 
MgTa$_2$N$_3$, which has already been synthesized \cite{MgTa2N3}, crystallizes in a hexagonal crystal structure with space-group $P6_3/mcm$ ($D_{6h}^3$, No.~193) as shown in Fig.~\ref{fig1_stru}(a). In this structure, Mg/Ta layers are sandwiched by N layers alternatively. Every Mg atom lies in the center of an octahedron of N atoms. Ta atoms are divided into two groups: $\delta$- and $\theta$-Ta. The $\delta$-Ta atoms, which are in the Mg layer, occupy the center of octahedra, similar to the case in $\delta$-TaN \cite{thetaTaN}, whereas $\theta$-Ta atoms lie in the adjacent plane and located in the center of pentahedra of N atoms, similar to that in $\theta$-TaN \cite{ScTaN2}. The calculated lattice constants are $a\!=\!5.218$ \AA~and $c\!=\!10.435$ \AA, consistent with the experimental values \cite{MgTa2N3}.

\begin{figure*}
\includegraphics[width =0.95\textwidth]{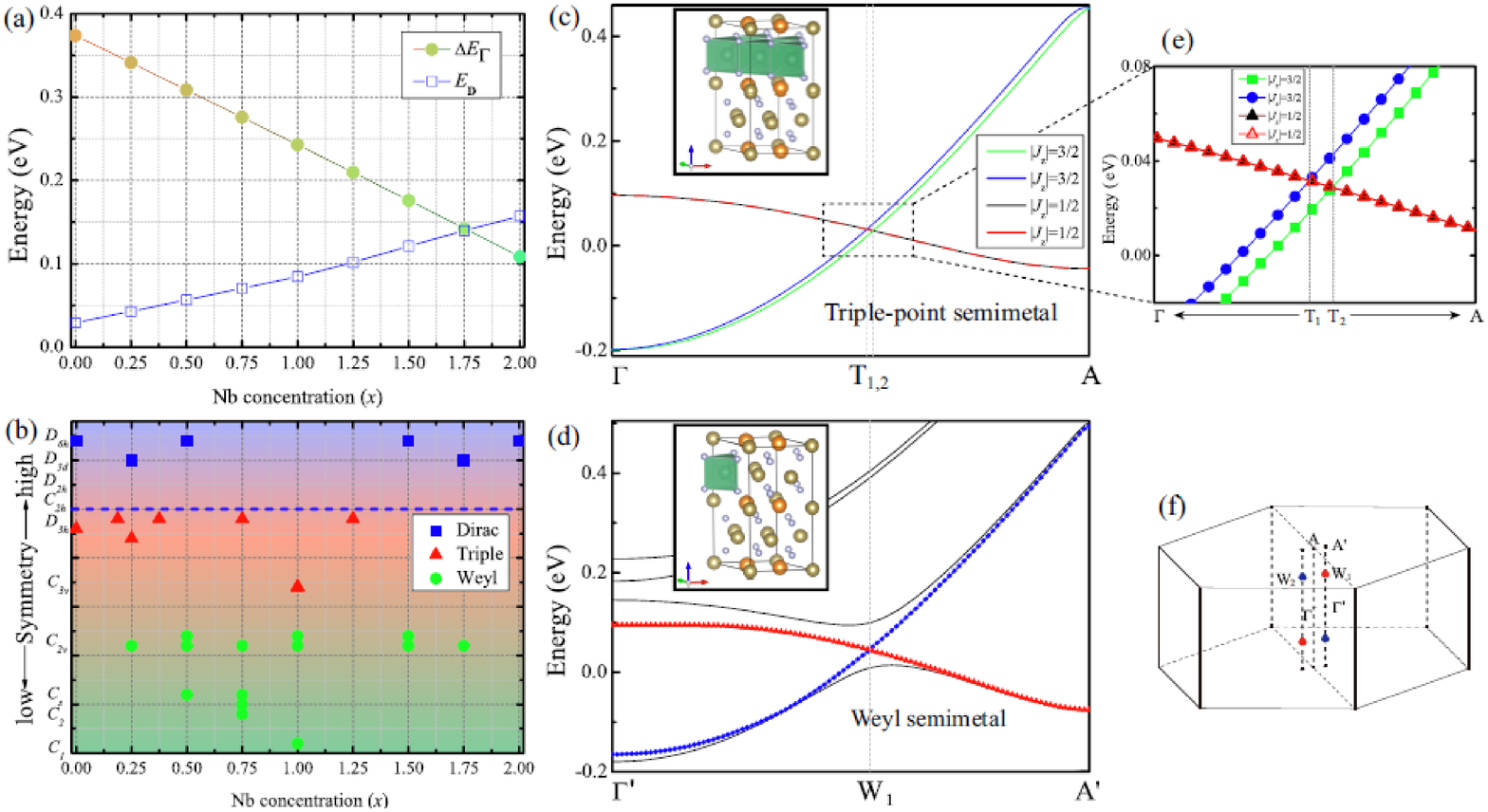}
\caption{\label{fig3_triple} (Color online) (a) Energy splitting between $\Gamma_8^+$ and $\Gamma_9^+$ states ($\Delta E_\Gamma$) and the energy of Dirac point ($E_D$) in MgTa$_{2-x}$Nb$_x$N$_3$ under the VCA. (b) Symmetry breaking effect of Nb doping using the supercell simulation. Each point represents a configuration with different symmetry. Band structures of example configurations that host (c) triple-point [MgTa$_{1.25}$Nb$_{0.75}$N$_3$($D_{3h}$)] and (d) Weyl semimetal states [MgTa$_{1.75}$Nb$_{0.25}$N$_3$ ($C_{2v}$)]. The insets shows the corresponding atomic configurations. (e) the zoomin plot of band structures around the triple points. (f) Schematic illustration of the Weyl point distribution in the 3D Brillouin zone where red and blue dots represent Weyl points with Chern number of -1 or +1.}
\end{figure*}

Figure~\ref{fig1_stru}(c) shows the band structure of MgTa$_2$N$_3$ without considering spin-orbit coupling (SOC). Apparently, MgTa$_2$N$_3$ is a metal with both electron and hole pockets at the Fermi level, and there is a band crossing along $\Gamma$-$A$ due to the band inversion between the $B_{1u}$ and $E_{2g}$ states at $\Gamma$. The orbital-resolved bands clearly show that the crossing bands consist of one band composed of Ta $d_{z^2}$ orbital and one double-degenerate band derived from $d_{x^2-y^2}$ and $d_{xy}$ orbitals. Because these two crossing bands belong to different irreducible representations of the $C_{3v}$ symmetry group of $\Gamma$-$A$, the band crossing point is threefold (or sixfold if the spin degree of freedom is considered) degenerate and protected by the $C_{3v}$ rotational symmetry. This would suggest MgTa$_2$N$_3$ be a triple-point semimetal without considering SOC. However, since Ta is a heavy element with strong SOC, we further calculated the band structure in the presence of SOC as shown in Fig.~\ref{fig1_stru}(d). The $d_{x^2-y^2}$ and $d_{xy}$ dominated $E_{2g}$ state splits into $\Gamma_8^+$ and $\Gamma_9^+$ states with $J_z\!=\!\pm 1/2$ and $\pm3/2$ respectively ($J_z$ is the total angular momentum), while the $d_{z^2}$ derived $B_{1u}$ state becomes the $\Gamma_8^-$ state with $J_z\!=\!\pm1/2$. Since both time-reversal and inversion symmetries are present, every band is spin degenerate. Duo to the $C_{3v}$ symmetry, the $\Lambda_8$ and $\Lambda_9$ bands with different irreducible representations cannot hybridize with each other \footnotemark[\value{footnote}]. This leads to a symmetry-protected fourfold-degenerate band-crossing point along $\Gamma$-$A$. Hence the system is actually a three-dimensional (3D) Dirac semimetal with a pair of Dirac point at $(0,0,k_z^D\!=\!\pm0.293)$ (in unit of $2\pi/c$) with an energy of $E_D=29$ meV above the Fermi level due to the small electron pocket around $M$. We also calculated the Fermi velocity of the Dirac cone ($v_x\!=\!2.07$, $v_y\!=\!2.08$, $v_z^{+/-}\!=\!2.87/1.11$) eV\AA~which is comparable with that of the typical Dirac semimetal Na$_3$Bi \cite{Na3Bi_exp1}. Different from Na$_3$Bi, the Dirac cone in MgTa$_2$N$_3$ is more isotropic but tilt slightly along $k_z$ direction. Moreover, the Dirac cone is mainly composed of \textit{d} orbitals pertaining to the investigation of strong correlation effect or magnetism in \textit{d}-electron-mediated Dirac fermion systems \cite{PhysRevLett.109.206802}.

Subsequently, we studied the alloying effect in MgTa$_{2-x}$Nb$_x$N$_3$ by the virtual crystal approximation (VCA) \cite{adachi2009properties}. The VCA treatment typically gives a reasonable description for solid-solution systems in which the dopant and host atoms have a similar chemical character. As shown in Fig.~\ref{fig3_triple}(a), with the varying concentration of Nb, the energy splitting between $\Gamma_8^+$ and $\Gamma_9^+$ decreases linearly. This is because the SOC strength of Nb is much weaker so that the SOC-induced splitting  becomes smaller with the increasing Nb alloying concentration. Meanwhile, the band crossing points gradually move away from the Fermi level. We also calculated the topological invariants for different concentrations. Although Nb alloying changes the electronic structures around the Fermi level, the nontrivial band topology remains, as discussed later.

As the VCA method only gives an average effect without accounting for local inhomogeneity, we further adopted supercell simulations to investigate the effect of symmetry breaking induced by different local atomic environment in the alloyed MgTa$_{2-x}$Nb$_x$N$_3$. We studied about 40 configurations with different symmetry groups, alloying concentrations and supercell sizes (see Supplemental Material \footnotemark[\value{footnote}]). The results are presented in Fig.~\ref{fig3_triple}(b). Interestingly, we found that by breaking inversion symmetry while keep some crystalline symmetries (e.g., $C_{3v}$), a fourfold Dirac point splits into a pair of triply degenerate points along the $k_z$ axis. Further reducing the crystalline symmetries, the triple points becomes Weyl points departing from the $k_z$ axis. As examples of triple-point and Weyl semimetals, below we presented detailed results for two configurations with different symmetries and alloying concentrations: MgTa$_{1.25}$Nb$_{0.75}$N$_3$ ($D_{3h}$) and MgTa$_{1.75}$Nb$_{0.25}$N$_3$ ($C_{2v}$).

The band structures of the triple-point semimetal are shown in Fig.~\ref{fig3_triple}(c) and (e). Because the $D_{3h}$ crystalline symmetry of the alloyed system includes $C_{3v}$ but excludes inversion symmetry, two $|J_z|\!=\!3/2$ bands split gradually along $\Gamma$-$A$ and cross with the doubly degenerate $|J_z|\!=\!1/2$ bands, forming two pairs of triply degenerate points at $(0,0,k_z^{T_1}\!=\!\pm0.267)$ and $(0,0,k_z^{T_2}\!=\!\pm0.273)$ (in unit of $2\pi/c$), respectively, with an energy difference of $\Delta E^T=2.86$ meV. Unlike other candidate semimetals with multiple pairs of triple points \cite{triple1,triple2,tripleNature}, this system exhibits a minimal number of triple points as required by time-reversal symmetry, which provides an ``ideal'' base system to study the properties solely induced by triple-point fermions \cite{wang2017prediction}. This selectively Nb-alloyed MgTa$_{2-x}$Nb$_x$N$_3$ is expected to be grown by the molecular beam epitaxy techniques, which have been proved to be an atomically accurate growth method for high-quality samples, such as modulation-doped Ga$_{1-x}$Al$_x$As/GaAs superlattices \cite{stormer1979two,dingle1978electron}, doped III-V nitrides \cite{ng1998mbe,novikov2017molecular} and topological-insulator alloys (Bi$_{1-x}$Sb$_x$)$_2$Te$_3$ \cite{zhang2011band}. The experimental feasibility of synthesizing and growing the MgTa$_{2-x}$Nb$_x$N$_3$ alloys is discussed in detail in the Supplemental Materials \footnotemark[\value{footnote}]. Interestingly, we also found that breaking symmetries by substitutional alloying cannot drive the $p$-electron Dirac fermion in Na$_3$Bi into triple-point fermions \cite{Na3BiSb}\footnotemark[\value{footnote}], indicating that our finding might be a unique feature of $d$-electron-media Dirac fermions, such as the case in MgTa$_2$N$_3$.

The Weyl semimetal state is obtained by further reducing the crystalline symmetry. As shown in Fig.~\ref{fig3_triple}(d), two nondegenerate bands cross each other along $\Gamma^\prime$-$A^\prime$ which is slight away from the $\Gamma$-$A$ line. This leads to the formation of Weyl points at $(0.00779, -0.0003, \pm0.269)$ and $(-0.00749,0.0003,\pm0.269)$ [in units of ($2\pi/a$,$2\pi/a$,$2\pi/c$)] that are related by time-reversal and $C_{2v}$ symmetries [see Fig.~\ref{fig3_triple}(f)]. The average Wannier charge centers are calculated by the Wilson-loop method on the sphere enclosing the Weyl point, from which the Chern number ($-1$/$+1$) of each Weyl points ($W_1/W_2$) are obtained (see Supplemental Material \footnotemark[\value{footnote}]). Note that this system exhibits only two pairs of Weyl points---the minimum number possible in time-reversal invariant systems, which provides an ``ideal'' platform for studying exotic properties of Weyl fermions in experiments.

\begin{figure*}
\includegraphics[width =0.95\textwidth]{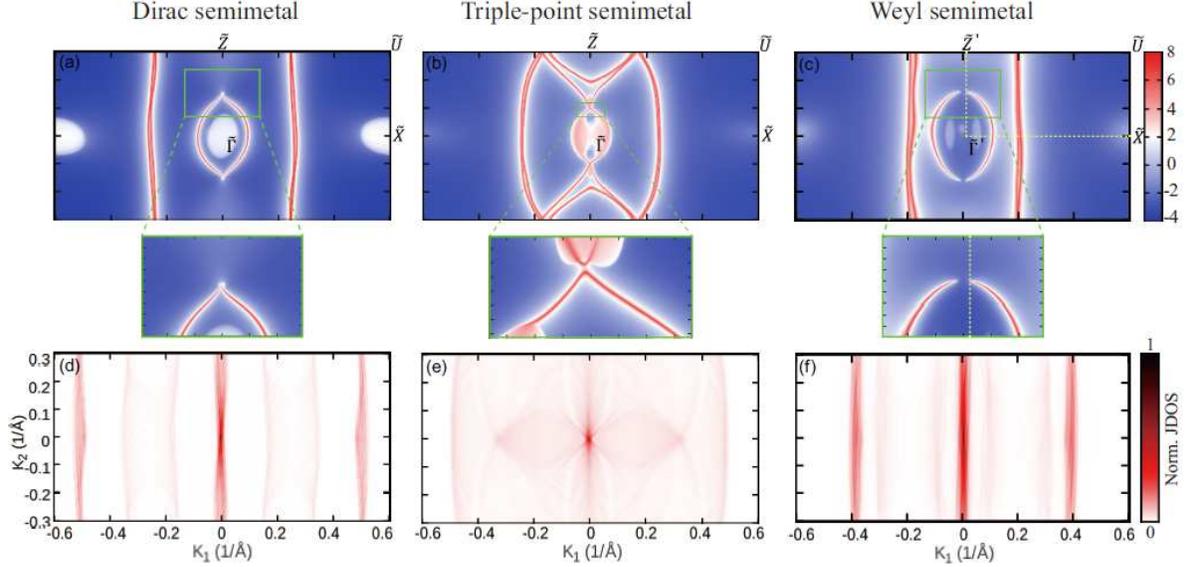}
\caption{\label{fig4_surf} (Color online) The projected isoenergy surface and QPI pattern for the (100) surface of MgTa$_{2-x}$Nb$_x$N$_3$. (a) and (d) The Fermi arc and QPI pattern at $E_D$ of the Dirac semimetal (MgTa$_2$N$_3$). (b) and (e) The Fermi surface and QPI pattern around the band crossings of the triple-point semimetal [MgTa$_{1.25}$Nb$_{0.75}$N$_3$ ($D_{3h}$)]. (c) and (f) The Fermi arc and QPI pattern at the energy of Weyl point $E_W$ of the Weyl semimetal [MgTa$_{1.75}$Nb$_{0.25}$N$_3$ ($C_{2v}$)]. The QPI patterns are derived from the surface-state-based calculations.}
\end{figure*}

To further reveal the transition from Dirac to triple-point fermions in the alloyed MgTa$_{2-x}$Nb$_x$N$_3$, we constructed a low-energy effective Hamiltonian using the theory of invariant \cite{bir1974symmetry}. 
Due to the $C_3$ rotation and inversion symmetries of the system, it is more convenient to make a linear combination of the Ta $d_{z^2}$, $d_{x^2-y^2}$ and $d_{xy}$ orbitals as $d_{\pm2}=d_{x^2-y^2}\pm i d_{xy}$, which have the $z$-direction orbital angular momentum $L_z\!=\!\pm2$. By including SOC into the hybridized orbtial picture, the $d_{z^2}$ orbital contains two states with $J_z\!=\!\pm\frac{1}{2}$ which compose the $\Lambda_8$ band, while the two $d_{\pm2}$ orbitals become four states with $J_z\!=\!\pm\frac{1}{2}, \pm\frac{3}{2}$, which contribute to the two SOC-split bands.
Using $\Gamma_8^-$ and $\Gamma_9^+$ states as basis, we can then derive an effective $4\times4$ Hamiltonian by considering the time-reversal and $D_{6h}^3$ symmetries.
\begin{equation}
H_{\mathrm{eff}}(\mathbf{k})=\left(
\begin{array}{cccc}
\epsilon_1(\mathbf{k}) & 0 & Ak_+ & 0\\
0 & \epsilon_1(\mathbf{k}) & 0 & -Ak_-\\
Ak_- & 0 & \epsilon_2(\mathbf{k}) & D(\mathbf{k})\\
0 & -Ak_+ & D^*(\mathbf{k}) & \epsilon_2(\mathbf{k})\\
\end{array}
\right),\nonumber
\end{equation}
where $\epsilon_j(\mathbf{k})=E_j+F_jk_z^2+G_j(k_x^2+k_y^2)$ ($j=1,2$) and $k_{\pm}=k_x\pm ik_y$. The term $D(\mathbf{k})$ describes the breaking of inversion symmetry, which should be zero for intrinsic MgTa$_2$N$_3$. The material-dependent parameters in the above Hamiltonian are determined by fitting the energy spectrum of the effective Hamiltonian to that of first-principles calculations. Defining $M(\mathbf{k})=\epsilon_1(\mathbf{k})-\epsilon_2(\mathbf{k})=M_0-M_1k_z^2-M_2(k_x^2+k_y^2)$, we obtained the eigenvalues of the above Hamiltonian $E(\mathbf{k})=\frac{1}{2}[\epsilon_1(\mathbf{k})+\epsilon_2(\mathbf{k})\pm\sqrt{M(\mathbf{k})^2+4 A^2k_+k_-}]$ and found two gapless solutions at $\mathbf{k}^D=(0,0,k_z^D=\pm\sqrt{\frac{M_0}{M_1}})$, which are nothing but the Dirac points discussed above. For the triple-point semimetal state in the alloyed MgTa$_{2-x}$Nb$_x$N$_3$ with $D_{3h}$ symmetry, the inversion breaking term $D(\mathbf{k})=D_1k_z$ is no longer vanishing. It is straightforward to derive that the triple points are  $\mathbf{k}^T=(0,0,k_z^T=\pm\sqrt{\frac{M_0}{M_1}+\frac{|D_1|^2}{4M_1^2}}\mp\frac{|D_1|}{2M_1}\approx \pm k_z^D\pm \frac{|D_1|}{2M_1})$, indicating that each Dirac point splits into a pair of triple points in the $k_z$ axis.

To confirm the nontrivial topological nature of MgTa$_2$N$_3$, we calculated the $\mathbb{Z}_2$ topological invariants by directly tracing the evolution of 1D hybrid Wannier charge centers \cite{wannier1,*wannier2} during a ``time-reversal pumping'' process as proposed by Soluyanov and Vanderbilt \cite{alexey2}. Although MgTa$_2$N$_3$ is a semimetal without a global gap, the electronic structure within the $k_z\!=\!0$ and $k_z\!=\!\pi$ planes are fully gapped, hence $\mathbb{Z}_2$ topological invariants are well defined in these planes and can be used to identify the band topology. Opposite to the case of Na$_3$Bi \cite{Na3Bi}, it is found that the $\mathbb{Z}_2\!=\!1$ for the $k_z\!=\!0$ plane, whereas $\mathbb{Z}_2\!=\!0$ for the $k_z\!=\!\pi$ plane in MgTa$_2$N$_3$. Therefore, a band inversion between the $\Lambda_9$ and $\Lambda_8$ bands should occur along the $k_z$ direction as seen above [see Fig.~\ref{fig1_stru}(d)]. Consequently, MgTa$_2$N$_3$ is a topologically nontrivial 3D Dirac semimetal with topological surface states and Fermi arcs appearing on its side surfaces. It is worth noting that the above calculation of topological invariant is also valid for the alloyed MgTa$_{2-x}$Nb$_x$N$_3$, which indicates that topological surface states also exist on the surfaces of triple-point and Weyl semimetal phases of the MgTa$_{2-x}$Nb$_x$N$_3$ alloy. We calculated the $\mathbb{Z}_2$ topological invariants of two representative configurations discussed above, confirming the nontrivial topological nature of triple-point and Weyl semimetal phases of MgTa$_{2-x}$Nb$_x$N$_3$. (see Supplemental Materials \footnotemark[\value{footnote}]).

In Fig.~\ref{fig4_surf}, we present the (100) surface states of MgTa$_{2-x}$Nb$_x$N$_3$ that host Dirac, triple and Weyl semimetal states, respectively. Since the $k_z\!=\!0$ and $k_z\!=\!\pi$ planes have different $\mathbb{Z}_2$ invariants for all topological semimetals in the MgTa$_{2-x}$Nb$_x$N$_3$ alloy, the number of surface states is odd and even along the $\tilde{\Gamma}$-$\tilde{X}$ and $\tilde{Z}$-$\tilde{U}$ lines of the surface BZ, respectively, for all cases in Fig.~\ref{fig4_surf}. For the Dirac semimetal, the surface state on the $k_z$ axis is twofold degenerate \footnotemark[\value{footnote}]. Hence two pieces of Fermi arcs stick together at two singularity points where the surface projections of bulk Dirac points appear, as shown in Fig.~\ref{fig4_surf}(a). In the triple-point semimetal, the Dirac point splits into two triple points along the $k_z$ axis, each of which contributes a nondegenerate surface state. Hence the Fermi arcs are gapped in the $k_z$ axis and an extra Fermi arc crosses the boundary of surface Brillouin zone to connect two Fermi pockets containing the projected triple points [see Fig.~\ref{fig4_surf}(b)]. In the Weyl semimetal, because a Dirac point splits into a pair of Weyl points with opposite Chern numbers, two Fermi arcs connecting the projected Weyl points split also and locate in the opposite sides of the $k_z$ axis [see Fig.~\ref{fig4_surf}(c)]. Having the Fermi arcs of different topological semimtals which can be measured by angle-resolved photoemission spectroscopy techniques, we further analyzed their fingerprints in QPI spectra\footnotemark[\value{footnote}], which can be observed by scanning tunneling spectroscopy (STS) measurements \cite{TCIQPI,DiracQPI,inoue2016quasiparticle,*batabyal2016visualizing,*zheng2016atomic,PhysRevLett.116.066601}. As show in Fig.~\ref{fig4_surf}(d)-(f), both Dirac and Weyl semimetals exhibit a stripe-type STS pattern, whereas the triple-point semimetal shows a flower-shape STS pattern with four petals due to the extra Fermi arc across the boundary of surface Brillouin zone. The QPI spectra reveal the scattering processes of surface states by defects, which are broadly relevant to surface transport and devices application based on these topological semimetals.

In conclusion, we significantly expand the horizon of a classical materials engineering approach, alloy engineering, to a new territory for realizing different fermions in one single material system. This is exemplified by calculating the phase transitions among Dirac, triple-point and Weyl semimetal states in the MgTa$_{2-x}$Nb$_x$N$_3$ system. Our finding provides useful guidance for designing topological semimetallic materials that can host different fermionic quasiparticles. Our proposed approach of engineering topological states by alloying is general, not only applicable to other 3D topological fermions but also extendable to 2D topological surface fermions \cite{PhysRevLett.119.196403,pariari2017coexistence} via surface alloying \cite{hosmani2014introduction,schmid2000alloying,lu2013order}.

\begin{acknowledgments}
This work was supported by U.S. DOE-BES (Grant No. DE-FG02-04ER46148). The calculations were done on the CHPC at the University of Utah and DOE-NERSC.
\end{acknowledgments}

\providecommand{\noopsort}[1]{}\providecommand{\singleletter}[1]{#1}%

\end{document}